\documentstyle[aps,amssymb,eqsecnum,epsf]{revtex}
\def \be{\begin{equation}}
\def \ee{\end{equation}}
\def \ba{\begin{eqnarray}}
\def \ea{\end{eqnarray}}

\begin{document}

\title{A cure for unstable numerical evolutions of single black holes:
adjusting the standard ADM equations}
\author{Bernard Kelly, Pablo Laguna, Keith Lockitch, Jorge Pullin, Erik Schnetter,
Deirdre Shoemaker and Manuel Tiglio}
\address{
Center for Gravitational Physics \& Geometry,\\
Penn State University, University Park, PA 16802}

\maketitle

\begin{abstract}
Numerical codes based on a direct implementation of 
the standard ADM formulation of Einstein's
equations have generally failed to provide long-term stable and convergent
evolutions of black hole spacetimes when excision is used to remove the
singularities.
We show that, for the case of a single black hole in spherical symmetry,
it is possible to circumvent these problems by adding to the evolution 
equations terms involving the constraints, thus adjusting the
standard ADM system.  
We investigate the effect that the choice of the lapse
and shift has on the stability properties of numerical simulations and thus
on the form of the added constraint term.
To facilitate this task, we introduce the concept of quasi well-posedness,
a version of well-posedness suitable for ADM-like systems involving second-order spatial
derivatives.
\end{abstract}

\section{Introduction}
\label{sec:intro}

In numerical relativity, Einstein's equations are usually
solved as an initial value problem. That is, the spacetime
is foliated with spacelike hypersurfaces. These hypersurfaces,
or slices, are characterized by their intrinsic geometry (spatial 
metric $g_{ij}$) and extrinsic curvature $K_{ij}$.
Subsequent slices in a given foliation are
connected via the lapse function $\alpha$ and
shift vector $\beta^i$ \cite{york_77}.
Under this framework, the components of Einstein's equation 
naturally separate into constraint and evolution equations for the
dynamical variables $g_{ij}$ and $K_{ij}$.
Thus, a typical procedure to construct a spacetime
consists of first specifying
Cauchy data $(g_{ij},K_{ij})$ that satisfy the constraints
on the initial slice and then applying the evolution
equations to update these data into the next slice.
This procedure is known as free or unconstrained evolution.

In order to propagate $(g_{ij},K_{ij})$ using the evolution equations, 
one must provide in addition a prescription for choosing the
lapse function and shift vector.
One generally
regards the lapse-shift prescription as a choice of coordinates.
At the continuum level, given a lapse-shift choice and a specific set of
Cauchy data,
the evolution equations yield not only a unique 
spacetime metric expressed in a
given coordinate system but also
evolved data $(g_{ij},K_{ij})$ that
continues to satisfy the constraints.

Numerical approximations are likely to complicate the
picture described above.
For instance,
because of truncation errors, the data $(g_{ij},K_{ij})$ do
not satisfy the continuum Einstein constraints but rather their
discrete approximations.
Even if consistency between the numerical approximations
of the evolution
and constraint equations is achieved, the evolved data
$(g_{ij},K_{ij})$ will at best satisfy the constraints
up to truncation errors. Of course, this is what one should expect.
However, the presence of numerical errors could
also trigger fast growing modes that
render the numerical evolution unstable.
Characterizing and controlling these growing modes has been
and continues to be one of the most difficult
and demanding tasks in numerical relativity.

The origin of these destabilizing modes is multiple.
Unstable modes could be purely
numerical (e.g., the Courant instability) or they
could be present already at the continuum level.
An example of continuum instabilities are the so called
constraint-violating modes \cite{scheel}.
These are solutions of the Einstein evolution
equations that do not satisfy the Hamiltonian and
momentum constraints. The general perception is that
this class of solutions consists mostly of rapidly
growing solutions, although there is no formal
proof that this is indeed the case. In summary, even if one
were to start an evolution with ``perfect" initial data
that satisfies the constraints,
truncation errors could cause the numerical solution to drift into
a constraint-violating, and perhaps unbound, solution.

Various groups have accumulated numerical evidence supporting as a
possible cause of instabilities
the particular 3+1 form used to recast Einstein's equation as an 
initial value problem.
These observations have partially motivated the development
of hyperbolic formulations of Einstein equations
\cite{helmut,bruhat,wp_fritreul,bona,york,hyper_rev}. 
Hyperbolic formulations have the advantage that mathematical tools
exist to prove existence and uniqueness of the solutions as well
as well-posedness, thus providing important information for the
implementation of stable discretization algorithms.  
If excision is used to handle the singularities, numerical simulations
of black hole
spacetimes require the imposition 
not only of outgoing, radiative boundary conditions far away from the holes but also
of conditions on the excision boundary that respect the propagation of 
physical modes to flow down into the hole. 
Because hyperbolic formulations also yield
information regarding the
characteristics of all the field variables in
the system, in principle this knowledge could be extremely
useful when applying these boundary conditions. 

Unfortunately, so far, none of the numerical efforts based on
hyperbolic formulations of Einstein equations have been able to
demonstrate their clear superiority. 
Currently, the 
formulation originally developed by Shibata and Nakamura
\cite{shibanakam},
and later re-introduced by Baugarte and Shapiro \cite{baumshap}, (BSSN)
seems to be the least prone to instabilities. Interestingly enough,
the BSSN formulation is not explicitly hyperbolic. Just recently,
a numerical implementation of the Einstein-Christoffel hyperbolic system
\cite{york} has been applied to single black hole spacetime and
produced three-dimensional evolutions with stability properties
comparable to those using the BSSN formulation \cite{scheel2}.
The common lore these days is, however, that the 
standard Arnowitt-Deser-Misner (ADM) \cite{adm} formulation
is the one which most easily suffers from instabilities
\cite{alcub,cadm,miller}.
On the other hand, the ADM formulation has the advantage of
containing a minimal number of equations, making it attractive
for numerical studies.

The main objective of this paper is to show that, given a
choice of lapse function and shift vector, it is
possible to obtain long-term stable and convergent simulations
using the standard ADM formulation if ``appropriate" terms involving the
constraints
are added to the evolution equations.  
Of course, here the key ingredient is to understand what constitutes an
appropriate choice of constraint terms. We show that at least in
spherical
symmetry, it is possible to obtain a definite prescription for
adding these constraint-terms that only depends on the choice
of gauge or coordinate conditions, namely the lapse and shift. 
The idea of adding constraint-terms to the evolution equations 
(adjusted ADM formulations)
is not new \cite{lambda}. Just recently, Yoneda and 
Shinkai \cite{hisaaki2} studied in full detail the propagation of
the constraints in the family of adjusted ADM systems.
 
In order to help identify what constitutes a suitable choice of constraint-terms,
we introduce a definition of quasi well-posedness.
The idea of quasi well-posedness is simply 
to make choices of lapse and shift such that, when
the ADM system is enlarged with the addition of new variables
(typically, but not
always, quantities related to the first spatial derivatives of the three
metric), one obtains a strongly hyperbolic system. 
Once a strongly hyperbolic system is obtained, one can make use of the
standard results that apply to those systems regarding the
existence, uniqueness and well-posedness of the solutions.
The essence of quasi well-posedness is then that the properties of the
strongly hyperbolic system are
to some degree inherited by the original ADM system.
The advantage of this procedure is that by looking at the strongly
hyperbolic system,
one is also able to gain insight about the
propagation of the constraints and characteristics of the original
system,
thus improving the chances of obtaining stable numerical
evolutions.

In Sec.~\ref{sec:qwp}, we introduce the concept of quasi well-posedness. 
The explicit form of the equations and initial data 
used in our numerical evolutions are described in Sec.~\ref{sec:gkdot}. 
The prescription to adjust the standard ADM equations is presented
in Sec.~\ref{sec:adjust}. We investigate in Sec.~\ref{sec:gmodes}
the implications of the presence of gauge modes.
Next, in Sec.~\ref{sec:stable1d}, we discuss the
quasi well-posedness properties of the adjusted ADM system under
three lapse and shift conditions.
For each of these conditions, we present numerical results showing the
stability properties of the simulations. We end in Sec.~\ref{sec:end}
with concluding remarks.

\section{Quasi Well-Posedness}
\label{sec:qwp}

To understand the general idea of
quasi well-posedness, let us consider the initial
value problem for the wave equation 
$\partial_{tt}\,\varphi= \partial_{xx}\,\varphi$, ignoring boundaries.
By adding a new variable, $\pi=\partial_t\,\varphi$, this equation takes
the form 
\begin{equation}
\partial_t\,u = P \,\partial_{xx}\,u + Q u \;,  \label{wave1} 
\end{equation}
where
$$
u =
\left( 
\begin{array}{c}
\varphi  \\ \pi 
\end{array}
\right) \; , \;\;\;\;
P =
\left(
\begin{array}{cc}
0 & 0 \\
1 & 0 
\end{array} \right) \;\;\hbox{and}\;\;\;
Q =
\left(
\begin{array}{cc}
0 & 1 \\
0 & 0
\end{array} \right)\; .
$$
One can view the system of equations (\ref{wave1}) 
as the analogue of the standard ADM equations.
Now, adding one more variable, $\xi = \partial_x\,\varphi$, we rewrite
(\ref{wave1}) 
as a first order system in both time and space. That is,
\begin{equation}
\partial_t\,\tilde{u} = A\,\partial_x\, \tilde{u} + B \tilde{u} \; ,
\label{wave2}
\end{equation}
with 
$$
\tilde{u} =
\left( 
\begin{array}{c}
\varphi  \\ 
\pi  \\
\xi
\end{array}
\right) \; , \;\;\;
A =
\left(
\begin{array}{ccc}
0 & 0 & 0 \\
0 & 0 & 1 \\
0 & 1 &  0
\end{array} \right) \;\;\hbox{and}\;\;\;
B =
\left(
\begin{array}{ccc}
0 & 1 & 0\\
0 & 0 & 0 \\
0 & 0 & 0
\end{array} \right)  \;.
$$
Notice that one needs to supplement the system (\ref{wave2}) with the
the constraint $C= \xi -\partial_x\,\varphi$.
For this simple case, 
one can find estimates for the solution $u$ from 
(\ref{wave1}) directly, but the idea here is not to do so but rather to
use what is
known about the first order system (\ref{wave2}). The matrix $A$ has the
following set
of eigenvalues and eigenvectors:
\be
\begin{array}{lll}
\lambda_1 & =    1 \; &\mbox{ with eigenvector}  \;\;\; \vec{e}_1 =
[0,1,1]\\
\lambda_2 & =    0 \; &\mbox{ with eigenvector}  \;\;\; \vec{e}_2 =
[1,0,0]\\
\lambda_3 & =   -1 \; &\mbox{ with eigenvector}  \;\;\; \vec{e}_3 =
[0,-1,1]\,.
\end{array}
\ee
Given that the matrix $A$ has real eigenvalues and a 
complete system of eigenvectors, the system of equations
(\ref{wave2}) is strongly hyperbolic
\footnote{Following \cite{bertil}, a system is called symmetric
hyperbolic if the matrix $A$ is Hermitian. It is called strictly
hyperbolic
if the eigenvalues are real and distinct; it is called strongly
hyperbolic
if the eigenvalues are real and there exists a complete system of
eigenvectors;
and, finally, it is called weakly hyperbolic if the eigenvalues are
real.}
and thus well-posed. In this case, the 
matrix $A$ is also Hermitian (symmetric hyperbolicity).
and has real and distinct eigenvalues (strictly hyperbolic)

Since the system (\ref{wave2}) is well-posed, we can establish existence
and uniqueness of the solution to (\ref{wave2}), together with the usual
condition:
\begin{equation}
||\tilde{u}(t,\;)||^2 \leq D\,e^{a\, t} ||\tilde{u}(0,\;)||^2\;,
 \label{estimate} 
\end{equation}
with $a$ and $D$ independent of the initial data and $||\;||$ the
$L_2$ norm.
Furthermore, because the vector $\tilde{u}$ contains $u$ as part of its
components, 
$$
||u(t,\;)||^2 \leq ||\tilde{u}(t,\;)||^2 \;.  
$$
It follows then from equation (\ref{estimate}) that the solution of the
system (\ref{wave1}) satisfies the condition 
\begin{equation}
||u(t,\;)||^2 \leq D\,e^{a\, t}\left( ||u(0,\;)||^2 +
||\partial_x\,u(0,\;)||^2 \right)  \; .  \label{qwp}
\end{equation}
We should remark that, although (\ref{qwp}) was derived using $L_2$
norms,
the solution could in general be bounded by the initial data in some
other Sobolev norm.
The importance of the condition (\ref{qwp}) and similar bounds is that
it guarantees, among other things,  that the solutions 
to (\ref{wave1}) will not have unbounded high frequency modes.
In numerical calculations however, 
this is not enough. For practical purposes, even large power-law growths such as
$t^4$ are likely to be extremely problematic \cite{scheel}.

The idea is then to investigate quasi well-posedness in the standard ADM 
formulation. That is, given a choice of lapse and shift, one enlarges
the 
standard ADM system introducing new variables that render the system at 
least strongly hyperbolic. Existence, uniqueness and well-posedness for
the 
ADM equations will then follow from the corresponding usual properties
of 
the strongly hyperbolic system.  In some instances, to obtain such
hyperbolic 
formulations it is not enough just to add the spatial derivatives of the 
metric as new variables, but one also has to add the constraints to the 
evolution equations, or even take combinations of these evolution
equations. 
But the general idea is the same: add the constraints and/or take the
same 
combination in the ADM equations as one does in the strongly hyperbolic
system.  
It will be in this sense that we will refer, in what follows, to the
quasi 
well-posedness of some ADM systems.

It is important to mention that for this analysis we have neglected the 
effects from boundary conditions. For the case under consideration
(single black 
hole evolutions via excision of the singularity) boundary conditions at 
the excised region do not affect quasi well-posedness since, as we shall 
see, all the characteristics of the system are outgoing into the hole. 
At the outer edge of the computational domain, we use the exact analytic solution 
as boundary condition.  Our numerical results 
show that this does not influence the properties drawn from quasi 
well-posedness for the class 
of gauge choices and added constraint-terms we considered.

\section{The $\dot g$ -- $\dot K$ Equations}
\label{sec:gkdot}

Although the idea of quasi well-posedness 
can be applied to more general systems, throughout this paper
we will concentrate on the ADM evolution equations ($\dot g - \dot K$
equations)
obtained from the vanishing of the Ricci tensor, i.e. 
\begin{eqnarray}
\partial_t g_{ij}-{\cal L}_{\beta} g_{ij} &=& - 2\, \alpha\, K_{ij} \, ,
\label{gdot} \\ 
\partial_t K_{ij}-{\cal L}_{\beta} K_{ij} &=&  - \nabla_i
\nabla_j \alpha +   \alpha\, ( R_{ij} + K\, K_{ij} - 2 K_{ik}\, K^k\,_j
) \label{Kdot}
\end{eqnarray}
As customary in numerical relativity,
we will focus on free evolutions; that is, we will not enforce the
constraints
\begin{eqnarray}
2\,H &\equiv & R + K^2 - K_{ij} K^{ij} = 0  \label{hamcons}\\
M^i & \equiv & \nabla_j (K^{ij} - g^{ij} K )  = 0 \, ,\label{momcons}
\end{eqnarray}
except at the initial data.
Finally, to facilitate our analysis, we will only consider a single
black
hole in spherical coordinates.
The most general time-dependent spherically symmetric metric in a 3+1
form is:
\begin{eqnarray}
ds^2 = (-\alpha^2 + a^2\,\beta^2)\, dt^2 
+ 2\,a^2\,\beta\, dt\, dr + a^2\, dr^2 + b^2\, d \Omega^2\,,
\label{spherical}
\end{eqnarray}
where all functions are assumed to depend only on $r$ and $t$;
$d\Omega^2  \equiv d\theta^2+\sin^2\theta\,d\phi^2$, and $\beta  \equiv
\beta^r$. In these spherical coordinates, the spatial metric,
extrinsic curvature and Ricci tensor are diagonal:
\begin{eqnarray}
g_{ij} &=& \hbox{diag}(a^2,\, b^2,\,
b^2\,\sin^2\theta)\label{eq:3metric}\\
K^i\,_j &=& \hbox{diag}(K_a,\, K_b,\, K_b) \label{eq:kk}\\
R^i\,_j &=& \hbox{diag}(R_a,\, R_b,\, R_b)\,. \label{eq:ricci}
\end{eqnarray}
The $\dot g$--$\dot K$ equations take the form
\ba
\left(\partial_t - \beta\,\partial_r\right)\,a &=& -\alpha\,a\,K_a
                  + a\,\partial_r\,\beta
\label{adot}\\
\left(\partial_t - \beta\,\partial_r\right)\,b &=& -\alpha\,b\,K_b
\label{bdot} \\
\left(\partial_t - \beta\,\partial_r\right)\,K_a &=&
-\frac{1}{a^2}\left(\partial_r^2\alpha
- \frac{1}{a}\partial_r a\,\partial_r \alpha\right)
+ \alpha\,\left[R_a + \left(K_a + 2\,K_b\right)\,K_a\right]
\label{kadot} \\
\left(\partial_t - \beta\,\partial_r\right)\,K_b &=&
-\frac{1}{b\,a^2}
\partial_r b\,\partial_r \alpha
+ \alpha\,\left[R_b + \left(K_a + 2\,K_b\right)\,K_b\right]\,,
\label{kbdot}
\ea
and the constraints
\ba
H & \equiv &  \frac{R_a}{2} + R_b + 2\,K_a\,K_b + K_b^2 =
0\label{eq:h_const}\\
M & \equiv &  \partial_r K_b + (K_b-K_a)\,\frac{\partial_r b}{b} =
0\label{eq:m_const}\,,
\ea
where the components of the Ricci tensor are given by
\ba
R_a &=& \frac{2}{a^3\,b}\left(-a\,\partial_r^2b + \partial_r
a\,\partial_r b\right) \\
R_b &=& \frac{1}{a^3\,b^2}\left[-b\,a\,\partial_r^2b + b\,\partial_r
a\,\partial_r b
+a^3-a\,\left(\partial_r b\right)^2\right]\,.
\ea

Our numerical results consist of free-evolutions of analytic,
single black hole initial data $(a,\, b,\, K_a,\, K_b)$ in a
computational
domain $r_e \le r \le r_o$, where $r_o$ denotes the location
of the outer boundary far from the hole and $r_e$ the excision boundary 
inside the black hole horizon. 
The numerical code used
solves  the $\dot g - \dot K$ equations in the interior of the
computational domain by the method of lines. 
That is, the $\dot g-\dot K$ system has the generic form 
\be 
\partial_t\, u = \beta\,\partial_r\, u + \rho\,, \label{eq:u} 
\ee 
with $u = u(a,\, b,\, K_a,\, K_b)$ and
$\rho$ given by the r.h.s. of equations
(\ref{adot}--\ref{kbdot}).
The starting point
is then to approximate the
derivatives appearing in $\rho$ by
second order accurate, centered finite differences.  
On the other hand, the the spatial derivative in the
``advection'' term $\beta\,\partial_ru$ is
approximated by an upwind discretization \cite{fletcher} of the form:
\be
\partial_r\, u_i \approx \Delta_r u_i \equiv 
\frac{(u_{i+1}-u_{i-1})}{2\,\Delta r}
\pm q \frac{(u_{i\mp 1}-3\,u_{i}+3\,u_{i\pm 1}-u_{i\pm 2})}{3\,\Delta r}
\label{eq:advect}
\ee
where $q \ge 0$ and 
the choice of sign is given by $\pm = \hbox{sign}(\beta)$. As we shall
later see,
for the black hole spacetime metrics under consideration, the shift
vector is always non-negative. Thus, the discretization
(\ref{eq:advect}) is only needed with the upper sign.
The above upwind discretization has truncation error:
\be
\tau_i \equiv \Delta_r u_i - \partial_r\, u_i = 
\frac{1}{6}\,(1-2\,q)\,\Delta r^2\,\partial^3_r \,u_i
\mp \frac{q}{6}\,\Delta r^3\,\partial^4_r u_i\,.
\label{eq:truncation}
\ee
Thus, the discretization is second order accurate for any choice of the
adjustable parameter $q$, except for $q=0.5$ when the truncation errors
are of $O(\Delta r^3)$. In our simulations, we set $q=0.5$. This value yields both stable
evolutions and minimizes the amount of dissipation introduced by the discretization. 
After the discretization of the spatial differential 
operators, Eq.~(\ref{eq:u}) becomes a
set of ordinary differential equations for the interior grid points
$\lbrace r_i \rbrace_{i=1,...,N-1}$. The grid values
$r_0$ and $r_N$ denote the locations of the inner and outer boundaries,
respectively.
The temporal updating of these equations is carried out via an iterated
Crank Nicholson method \cite{saul}. 
Notice that at the outer-most grid point $r_{N-1}$, 
the upwind scheme (\ref{eq:advect}) requires ``ghost" values
at $r_{N+1}$. These values are computed using second or third
order accurate extrapolations. 

Next is the implementation of boundary conditions.
As mentioned before, there are two boundaries in the problem,
one far from the black hole at $r_N = r_o$ and 
a second at the point of excision, $r_0 = r_e$.
The boundary condition used at $r_o$ is that 
all field variables take
the values provided by the exact analytic solution. 
It is important to notice that in previous efforts in numerical
evolutions of 
spherically symmetric black hole spacetimes, the stability of the
evolutions
depended not only in the location of the outer boundary but also 
on the imposition of ``outgoing" boundary conditions. 
One of these outgoing conditions
is constructed by assuming that the
solution error $u - \bar u$ is of the form
$\epsilon(t-r)/r^n$, with $\bar u$ the analytic exact solution.
This condition yields
\be
\partial_t\,u = -\partial_r\,u + \partial_r\,\bar u + n\,\frac{(u-\bar
u)}{r} \,,
\label{eq:outgoing}
\ee
to be applied at the nodal grid point $r_N = r_o$
The outgoing condition (\ref{eq:outgoing}) is then added to the 
system (\ref{eq:u}) and handled with the same method of lines used for
the
interior points.
Obviously, there is no reason to believe
that the solution error should behave as $\epsilon(t-r)/r^n$ for general
choices of lapse and shift. Another approach to construct outer boundary 
conditions that has been used is to 
blend \cite{gomez} the numerical solution to the analytic one
beyond a certain radius, $r$.

Finally, at the inner or excision boundary, the working assumption is
that
all of the fields have outgoing (into the hole) characteristics.
We will show that for the gauge or coordinate conditions under
consideration,
this is indeed the situation.
Therefore, there is no need to impose any boundary condition,
and one can just apply at $r_e$ the same system of equations used
in the interior of the computational domain. However, finite difference
discretization would require ``ghost" values at $r_{-1}$.
We construct these values by extrapolation.
Another possibility, which is becoming popular in three-dimensional
simulations 
\cite{miguel}, is instead to extrapolate the r.h.s. of Eq.~(\ref{eq:u})
to $r_0 = r_e$. Either approach was stable for the cases considered 
in the present work.

An important requirement when performing 
numerical simulations of black
holes in which the singularities are excised from the computational domain
is to use coordinates that are regular and penetrate the horizon.
Foliations that penetrate the horizon facilitate the task of removing
(i.e. excising) a region containing the black hole singularity while
preserving the causal structure of the spacetime exterior to the event
horizon.  Thus the
numerical results we present here 
consist of numerical evolutions that for
infinite resolution correspond to the
solution of a single black hole 
expressed in ingoing-Eddington-Finkelstein (iEF) and also
Painlev\'e-Gullstrand (PG) \cite{PG1,PG2}.  coordinates.
The iEF coordinates coincide with Kerr-Schild coordinates
in the case of zero angular momentum \cite{mtw}. 
Recently, Martel and Poisson \cite{Poisson}
have showed that the PG and iEF coordinates
are members of the same one-parameter family of coordinate systems.

In iEF coordinates, the
line element (\ref{spherical}) takes the explicit form
\begin{equation}  ds^2 = -\left( 1-\frac {2\,m}{r}
\right) dt^2 + \frac {4m}{r} dt\, dr + \left   ( 1+\frac {2\,m}{r}
\right) dr^2 +
r^2 d\Omega ^2 \;.    \label{ef} 
\end{equation} 
The ADM variables in these coordinates are given by:
\ba
\alpha &=& \left(1+\frac{2\,m}{r}\right)^{-1/2} \label{alphaiEF} \\
\beta &=& \frac{2\,m}{r}\left(1+\frac{2\,m}{ r}\right)^{-1}
\label{betaiEF} \\
a &=& \left(1+\frac{2\,m}{r}\right)^{1/2}\label{aiEF} \\
b &=& r \label{biEF}\\
K_a &=&
-\frac{2\,m}{r^3}(r+m)\left(1+\frac{2\,m}{r}\right)^{-3/2}\label{kaiEF}\\
K_b &=&
\frac{2\,m}{r^2}\left(1+\frac{2\,m}{r}\right)^{-1/2}\,.\label{kbiEF}
\ea
Similarly, for PG coordinates
\begin{equation}  
ds^2 = - \left( 1-\frac {2\,m}{r}\right) dt^2 
+ 2 \sqrt{\frac{2\,m}{r}} dt\, dr 
+ dr^2 + r^2 d\Omega ^2 \;,    
\label{eq:PG} 
\end{equation} 
and
\ba
\alpha &=& 1 \label{alphaiPG} \\
\beta &=& \sqrt{\frac{2\,m}{r}} \label{betaiPG} \\
a &=& 1\label{aiPG} \\
b &=& r \label{biPG}\\
K_a &=& -\frac{\beta}{2\,r}\label{kaiPG}\\
K_b &=& \frac{\beta}{r}\,.\label{kbiPG}
\ea

The geometrical interpretation of the iEF coordinate system is that
in addition to having a
timelike killing vector, the combination of timelike and
radial tangent vectors $\vec\partial_t - \vec\partial_r$ remains null.
In terms of the spacetime metric, this condition is stated as
$g_{tt}-2\,g_{tr}+g_{rr} = 0$, or similarly in terms of 3+1 metric
functions as $\alpha = a\,(1-\beta)$.
On the other hand, the PG coordinate system can be viewed as that 
anchored to a family of freely moving observers (time-like) 
starting at infinity with vanishing velocity \cite{Poisson}.

Finally, in numerical evolutions of spherically symmetric spacetimes,
it is useful to monitor not only the Hamiltonian and momentum constraints
but to pay attention also to the mass function
\be
{\cal M}(r,t) = \frac{b}{2}(1- \nabla_{\mu}b\,\nabla^{\mu} b)\,.
\label{eq:mass_f}
\ee
In vacuum, this ${\cal M}$ is the gauge invariant definition of mass.

\section{Adjusted ADM Systems}
\label{sec:adjust}

Each of the equations in the $\dot g$--$\dot K$ or standard ADM system (\ref{adot}-\ref{kbdot})
has the form
\begin{equation}
\partial_t\,u_{(n)} - \beta \,\partial_r\,u_{(n)} =  
B_{(n)}\, u_{(n)} + C_{(n)}\label{eq:sadm}
\end{equation}
(no summation over $n$) with
\ba
u_{(n)} &=& (a,\, b,\, K_a,\, K_b)\\
B_{(n)} &=& \hbox{diag}(-\alpha\,K_a + \partial_r\,\beta,\, 
-\alpha\,K_b,\, 2\,\alpha\,K_b,\, \alpha\,K_a)
\ea
The $C_{(n)}$ contain the remaining terms that cannot be written in
the form $B_{(n)}\,u_{(n)}$, with $B_{(n)}$ independent of $u_{(n)}$.
If one views each equation as independent, that is
$\beta$, $B_{(n)}$ and $C_{(n)}$ as given and independent of $u_{(n)}$,
Eqs.~(\ref{eq:sadm}) admit rapidly growing solutions if $B_{(n)} > 0$,
exponentially growing if $B_{(n)}$ and $\beta$ are non-negative constants.
Obviously, when the equations (\ref{eq:sadm}) are considered as a 
coupled system of equations, one cannot guarantee the above conclusion.
However, we have been able to track the problems observed in our numerical evolutions
to those terms for which $B_{(n)} > 0$. 

In iEF and PG coordinates, the second and fourth components of
$B_{(n)}$ are non-positive. The first component of $B_{(n)}$ vanishes
for PG coordinates and is positive for iEF. Finally, the third component
of $B_{(n)}$ is always non-negative, for both iEF and PG coordinates.
Our numerical experiments indicate that the origin of the instabilities 
is due to the term involving $B_{(n=3)} = 2\,\alpha\,K_b$.
The term in iEF coordinates 
involving $B_{(n=1)}$ with the ``wrong" sign 
did not seem to affect the evolutions (see results below).

The objective is then to find a way to ``change the sign" of
$B_{(n=3)}$. Fortunately, the combination $K_b\,K_a$ also appears in the 
Hamiltonian constraint (\ref{eq:h_const}). One can then add to the evolution
equation for $K_a$ a term of the form $-\mu\,\alpha\,H$. As a consequence,
$B_{(n=3)} = 2\,(1-\mu)\,\alpha\,K_b$. Therefore, in principle, any value
of $\mu \ge 1$ should produce stable evolutions. Our numerical simulations 
show that $\mu=2$ is an optimal value to reach quickly the time independent solution.
However, the same experiments indicate also that even values of $\mu \approx 0.5$ 
yield stable evolutions. The reason for this is likely because the simple analysis
above does not take into consideration the non-linear coupling in the equations.

\section{Linear Gauge Modes}
\label{sec:gmodes}

A potential source of
instabilities in numerical simulations is the presence of gauge modes. 
Gauge modes arise because
a prescription of the lapse and shift is not sufficient for complete
gauge fixing.
If the lapse and shift functions are for instance
determined from algebraic expressions or differential equations, there is still
remaining freedom left to perform coordinate transformation that
take us from $(g_{ij},K_{ij})$ to $(\tilde g_{ij},\tilde K_{ij})$, leaving the
lapse-shift prescription invariant. In other words, the
data $(g_{ij},K_{ij})$ will be unique up to coordinate transformations
that leave the lapse-shift prescription invariant. One can then encounter the
situation in which the
transformed pair $(\tilde g_{ij},\tilde K_{ij})$ continues  to
satisfy the constraints, but it possesses unbounded growths.
In principle, because $(\tilde g_{ij},\tilde K_{ij})$
are allowed solutions to the evolution
equations, one could through numerical convergence and
monitoring of the constraints single out these modes.
In practice the situation is not that simple. Gauge conditions that allow
the rapid growth of grid functions are likely to trigger numerical instabilities.

To investigate these modes, let us consider an infinitesimal coordinate transformation,
\be
x^\mu \rightarrow x^\mu+\xi^\mu\,,
\ee
This transformation induces a linear perturbation of the spacetime metric,
\be
g_{\mu \nu} \rightarrow g_{\mu \nu} + \delta g_{\mu\nu}\,,
\label{gshift}
\ee
of the form
\be
 \delta g_{\mu\nu} = -{\cal L}_{\xi}\,g_{\mu \nu}.
\label{h_form}
\ee

In spherical symmetry, the most general coordinate transformation is
\ba
   t &\rightarrow&  t + \delta t(r,\,t) \label{gauget}\\
   r &\rightarrow& r + \delta r(r,\,t) \label{gauger}\,,
\ea
or equivalently
\be
  \xi^{\mu} = \left( \delta t, \delta r, 0, 0 \right)\,.
\label{xi}
\ee

Given (\ref{xi}), the gauge induced perturbations of the metric
takes the form
\ba
\delta g_{tt} &=& -2 \left(-\alpha^2+a^2\beta^2\right)\,\partial_t
\delta t
-2 a^2 \beta \, \partial_t \delta r
- \partial_t \left(-\alpha^2+a^2\beta^2\right)\,\delta t
- \partial_r \left(-\alpha^2+a^2\beta^2\right)\,\delta r
  \label{stt} \\
\delta g_{tr} &=& - a^2\beta\,\partial_t \delta t
- a^2 \, \partial_t \delta r
- \left(-\alpha^2+a^2\beta^2\right)\,\partial_r \delta t
- a^2\beta \, \partial_r \delta r
- \partial_t \left(a^2\beta\right)\,\delta t
- \partial_r \left(a^2\beta\right)\,\delta r
\label{str}  \\
\delta g_{rr} &=& -2 a^2\beta\,\partial_r \delta t
- 2 a^2\, \partial_r \delta r
- \partial_t \left(a^2\right)\,\delta t
- \partial_r \left(a^2\right)\,\delta r
\label{srr}  \\
\delta g_{\theta \theta} &=& - \partial_t \left(b^2\right)\,\delta t
- \partial_r \left(b^2\right)\,\delta r
\label{sthth} \\
\delta g_{\phi \phi} &=&
\delta g_{\theta\theta}\,\sin^2\theta \label{spp}\,,
\ea
with all remaining components vanishing.
On the other hand, from (\ref{spherical}), the 
components of the perturbed metric $\delta g_{\mu\nu}$ above are also given by:
\ba
\delta g_{tt} &=& -2\,\alpha\,\delta\alpha + 2\,a\,\delta a\,\beta^2
+2\,a^2\,\beta\,\delta\beta \label{eq:dgtt}\\
\delta g_{tr} &=& 2\,a\,\delta a\,\beta + a^2\,\delta\beta \label{eq:dgtr}\\
\delta g_{rr} & =& 2\,a\,\delta a\label{eq:dgrr}\\
\delta g_{\theta\theta} &=& 2\,b\,\delta b\label{eq:dgoo}\,.
\ea
In studying these gauge perturbations, we consider a number of different
prescriptions for choosing the lapse and shift.  These impose
restrictions on the components of $\delta g_{\mu\nu}$ 
to linear order, but are not 
sufficient for complete gauge fixing.

One piece of evidence usually given to argue for the highly unstable 
properties of the standard ADM formulation is its inability to simulate 
single black hole spacetimes.  Specifically, the problem consists of 
using a known analytic black hole solution of Einstein's equation to 
construct initial data, set boundary conditions and specify the lapse
and shift.  Given this input, the output is the numerical evolution of
the spatial metric $g_{ij}$ and extrinsic curvature $K_{ij}$.  
This simple setup does not yield stable evolutions.

Choosing the lapse and shift to be given by 
the exact analytic solutions implies that in the coordinate transformations
above, $\delta\alpha = \delta\beta = 0$.
We shall call this the ``exact lapse + exact shift" (EL+ES) condition, 
It is clear from (\ref{stt}--\ref{spp}) that the 
EL+ES choice does not fix the coordinates uniquely,
but rather results in a non-trivial system of PDE's to be satisfied by
$\delta t$ and $\delta r$.  The solutions to this PDE system form an
equivalence class of gauges all satisfying the same condition on the
lapse and shift.  At the analytic level, one selects a unique member
of this equivalence class by imposing boundary and initial conditions.
Numerically however, because of truncation errors, 
this is only possible if the evolution is stable.

Substitution of $\delta\alpha = \delta\beta = 0$ in
(\ref{eq:dgtt}-\ref{eq:dgrr}) yields
$\delta g_{tr} - \beta\,\delta g_{rr} =  \delta g_{tt}
- \beta^2\,\delta g_{rr} = 0$. 
Using (\ref{stt}--\ref{srr}), these conditions can be rewritten in terms of the gauge
perturbation $\xi^\mu$ as
\be
\partial_t\xi = A \partial_r\xi + B \xi
\ee
where 
$$
\xi =
\left(
\begin{array}{c}
\delta t  \\ \delta r
\end{array}
\right) \;\;\;\;\hbox{and}\;\;\;
A =
\left(
\begin{array}{cc}
\beta & 0 \\
\frac{\alpha^2}{a^2} & \beta
\end{array} \right)\,.
$$
The matrix $A$ has a degenerate eigenvalue $\lambda=\beta$ and 
corresponding eigenvector $\vec{e} = [0,\,1]$.  Therefore, the system is
only weakly
hyperbolic and thus ill-posed. It is then not possible to guarantee the 
absence of rapidly growing gauge modes. Gauge instabilities, by
themselves, 
do not violate the constraints; however, they are likely to trigger 
numerical instabilities and thereby couple to constraint-violating
modes.  
Note that this conclusion does not depend on the use of the standard ADM 
formulation, but it applies to any initial value formulation of Einstein's 
equation. This is likely the reason why it has not been possible to
produce hyperbolic formulations of Einstein equations using the
EL+ES lapse-shift condition.
Other groups have reported numerical instabilities associated 
with an EL+ES prescription. (See, for example, 
\cite{luis,BHS}.)
Our result, which provides some analytic insight into the source 
of these instabilities, extends the analysis of Ref.~\cite{luis} by  
including perturbations of a general spherically symmetric spacetime 
(\ref{spherical}) and by making no assumptions about the form of the 
coupling between gauge and constraint-violating perturbations.

Finally, the inability to guarantee the absence of unbound gauge modes does not
necessarily imply that it is impossible to design an evolution
scheme that is long-term stable and convergent. As we shall see,
by adjusting the standard ADM system with constraint terms,
stable evolutions are possible even in the presence of 
these gauge modes.

\section{Stable adjusted ADM systems in 1d}
\label{sec:stable1d}

We consider next a series of lapse-shift choices. For each choice, we investigate
(1) the quasi well-posedness of the resulting system of adjusted ADM evolution equations, 
(2) the propagation of the constraints and (3) the convergence
and stability of numerical simulations.

\subsection{Exact Lapse + Area Locking}

We can take advantage of the assumed spherical symmetry of the problem and 
``lock'' the area of constant-$r$ surfaces. That is, we exploit the lapse-shift freedom 
and set $\partial_t\,g_{\theta\theta} = 0$, or equivalently
$\partial_t\,{b} =0 \; \forall \; t$. From Eq.~(\ref{bdot}), this yields 
\begin{equation} 
0 = -\beta\, \partial_r\, b + \alpha\, b\, K_b \;\; ,   \label{lock}
\end{equation}
which can be seen as an algebraic equation to solve for $\beta$ or
$\alpha$. We will use (\ref{lock}) as
\begin{equation} 
\beta = \frac{\alpha\, b\, K_b}{\partial_r\,b} \;,   \label{elal_shift}
\end{equation}
with $b$ determined by the
initial data. In our case, for both iEF and PG coordinates, $b=r$. In addition, 
we choose an exact lapse, i.e. an arbitrary but {\it a
priori} specified function of spacetime. Here again, since the goal is to 
reproduce numerically the analytic solution, we set the lapse to
that given by the iEF or PG solutions. 

This exact lapse, area locking (EL+AL)
gauge condition has previously been investigated in Ref.~\cite{luis}. However,
the implementation of locking the area was done at the numerical level. That is,
condition (\ref{elal_shift}) was not explicitly used. Instead, during the 
temporal updating of grid functions,
a correction to the shift was introduced to keep the area locked. 
With this numerical area-locking and with blending outer boundary conditions, 
stable simulations were reported in Ref..~\cite{luis} 
for computational domains with $r_o \le 11\,m$.

Interestingly enough, this EL+AL choice of lapse and shift
yields an ADM system of equations already first order in space and quasi-linear; namely,
\begin{equation}
\partial_t\,u = A \,\partial_r\,u + B\, u \label{elal_evol}
\end{equation}
with 
$$
u =
\left( 
\begin{array}{c}
a \\
K_a  \\
K_b
\end{array}
\right) \;\;\;\hbox{and}\;\;\;
A =
\left(
\begin{array}{ccc}
\displaystyle{\beta } & 0 & 
\displaystyle{\alpha\, a\, r}  \\
& & \\
\displaystyle{\frac{r\,\partial_r\,\alpha  
+ 2\,(1-\mu)\,\alpha}{a^3\,r}}  & 
\displaystyle{\beta } & 0 \\
& & \\
\displaystyle{\frac{\alpha}{a^3\,r}} & 0 & 
\displaystyle{\beta } 
\end{array}
\right) \;\;,
$$
where $\beta $ is given by (\ref{elal_shift}) and we have set $b=r$ to
simplify notation. In the numerical evolutions, however, we do not
set $b=r$. The numerical code includes the evolution equation
for $b$. We do not explicitly write the
matrix $B$ since it is not necessary for the analysis below. However, we must emphasize that,
in order to achieve stability, 
a value of $\mu \ge 1$ is in principle needed (see Sec.~\ref{sec:adjust}). 

Although this EL+AL case is not
representative of the general ADM equations, where second spatial derivatives
do appear, we will use it to introduce the main techniques and ideas regarding
quasi well-posedness. 
We notice first that the matrix $A$ has eigenvalues
\begin{eqnarray}
\lambda _1 & = & \beta \label{elallam1} \\  
\lambda _2 & = & \beta + \frac{\alpha }{a} \label{elallam2} \\
\lambda _3 & = & \beta - \frac{\alpha }{a} \label{elallam3} \; ,
\end{eqnarray}
and corresponding eigenvectors
\ba
\vec{e}_1 &=& [0,1,0]\\
\vec{e}_2 &=& \left[ a^2\, r,
\frac{r\,\partial_r\,\alpha  + 2\,(1-\mu)\,\alpha}
{\alpha},1\right]\\
\vec{e}_3 &=& \left[-a^2\, r,
\frac{r\,\partial_r\,\alpha  + 2\,(1-\mu)\,\alpha}
{\alpha},1\right]\,.
\ea
Because all of the eigenvalues are real and distinct, 
the system (\ref{elal_evol}) is strictly hyperbolic independent of
the addition of the constraint term. 
The EL+AL system of equations is then an example
of a hyperbolic system that, unless
suitable constraint-terms are added, is subject to developing 
rapidly growing solutions (see Ref.~\cite{scheel} for another example). 
It is also important to notice that $\lambda_1$ represents
a characteristic speed corresponding to propagation along the timelike
normal to the foliation. Similarly, $\lambda_2$ and $\lambda_3$
are characteristic speeds along the light cone.

Let's now consider the particular case of initial data and lapse function
constructed from
the single black hole solution in iEF or PG coordinates.
The eigenvalues are given in these coordinates by 
\ba
\lambda_1 & = &\frac{2\,m}{r+2\,m} = \sqrt{\frac{2\,m}{r}}\\
\lambda_2 & = &1 = \sqrt{\frac{2\,m}{r}} + 1\\
\lambda_3 & = &\frac{2\,m-r}{2\,m+r} = \sqrt{\frac{2\,m}{r}} -1\,, 
\ea
where the first and second equalities are for iEF and PG coordinates, respectively.
Since the excision boundary $r_e$ is by construction inside the black hole
horizon (i.e. $r_e \le r_h \equiv 2\,m$), we have that all the eigenvalues
are positive there. By looking at the principal part of
Eq.~(\ref{elal_evol}), it is easy to see that non-negative eigenvalues
imply propagation of field variables in the direction of decreasing $r$-coordinate.
Therefore, at $r_e$ all the fields propagate out of the computational
domain into the hole singularity, thus boundary conditions are not
required. On the other hand, at the outer boundary $r_o$, one has $\lambda_3 < 0$,
with the remaining eigenvalues still non-negative. 
At $r_o$, one has then two ingoing modes
($\lambda_1$ and $\lambda_2$) 
and only one outgoing ($\lambda_3$). Given this information, it is perfectly possible to 
impose a condition suppressing modes entering the
computational domain. However, as we mentioned before, we choose not to do so
and impose at $r_o$ the analytic, exact solutions.
The direct consequence will be the appearance of a pulse at the outer boundary of the
computational domain due to discontinuities in the truncation errors
between the outermost evolved point and the boundary point $r_o$.
This pulse propagates in the direction of
the black hole and leaves the computational domain through the excision boundary.

The next step is to 
analyze the effect that the EL+AL choice has on the propagation of
the constraints. At the continuum level, 
for arbitrary choice of lapse, shift and initial
data ($g_{ij},\, K_{ij}$) 
satisfying the constraints (\ref{hamcons}) and (\ref{momcons}), 
the evolution equations (\ref{gdot}) and (\ref{Kdot}) guarantee, ignoring for the
moment boundary conditions, that the
evolved data will continue to satisfy the constraints. 
If one now takes into consideration boundary conditions,  
it is important to keep in mind that boundary data ($g_{ij},\, K_{ij}$)
must satisfy the constraints. By looking at the way the constraints propagate,
i.e. their characteristics, one gains insight about the allowed
boundary conditions consistent with the constraints.
Another important aspect of  
well-posedness in the propagation of the constraints is that
it guarantees that there will be no unbounded high 
frequency growth appearing in the constraints if
they are not exactly satisfied at the initial slice (for example, due to
numerical errors). This well-posedness for the constraint propagation is a 
non-trivial property, not possible to prove for generic 
formulation of Einstein's equations \cite{sfrittelli}.

What we look for are evolution equations for the constraints,
equations that would hold if the system (\ref{elal_evol}) is satisfied.
They can be found by taking time derivatives in both sides of equations
(\ref{eq:h_const}) and (\ref{eq:m_const}), replacing the time derivatives of the metric by
the r.h.s. of equations (\ref{adot}--\ref{kbdot}), and
finally expressing the metric and its spatial derivatives in terms of the
constraints and their spatial derivatives. Following this
procedure, it is not too difficult to show that 
\begin{equation}
\partial_t\,v = P\,\partial_r\, v + Q\, v \label{elal_cons} \; ,
\end{equation}
where now
\begin{equation}
v =
\left( 
\begin{array}{c}
H  \\ M
\end{array}
\right) \;\;\;\hbox{and}\;\;\;\;
P =
\left(
\begin{array}{cc}
\displaystyle{\beta } & 
\displaystyle{\frac{4\alpha }{a^2}}  \\
&  \\
\displaystyle{\frac{\alpha }{4}}  & 
\displaystyle{\beta } 
\end{array} \right) \; . \label{a_const}
\end{equation}
The matrix $P$ has eigenvalues 
$\bar\lambda _1 = \lambda_2 \; , \bar\lambda _2 = \lambda_3$ , with $\lambda _2$
and $\lambda _3$ given by (\ref{elallam2}) and (\ref{elallam3}), respectively. 
This implies that the system (\ref{elal_cons}) is also strictly hyperbolic
with characteristic speeds along the light cone. 
Notice that the constraints at $r_e$ propagate out of the computational
domain into the hole singularity, consistent with the outgoing propagation of
the field variables at $r_e$, namely the tilting, into the black hole,
of the light cone. At the outer boundary $r_o$, 
there is an ingoing mode ($\bar\lambda_2 >0$).
Therefore, one has, as expected, to be careful to provide data at $r_o$ consistent
with this entering mode. Since we are imposing at the outer boundary 
the analytic iEF and PG solutions, the data at $r_o$ already satisfy the constraints.
However, as mentioned before, choosing the lapse and shift does not completely
fix the gauge freedom, thus one still has to be careful handling
the gauge modes described in Sec.~\ref{sec:gmodes}.

\begin{figure}
\centerline{\epsfxsize=250pt\epsfbox{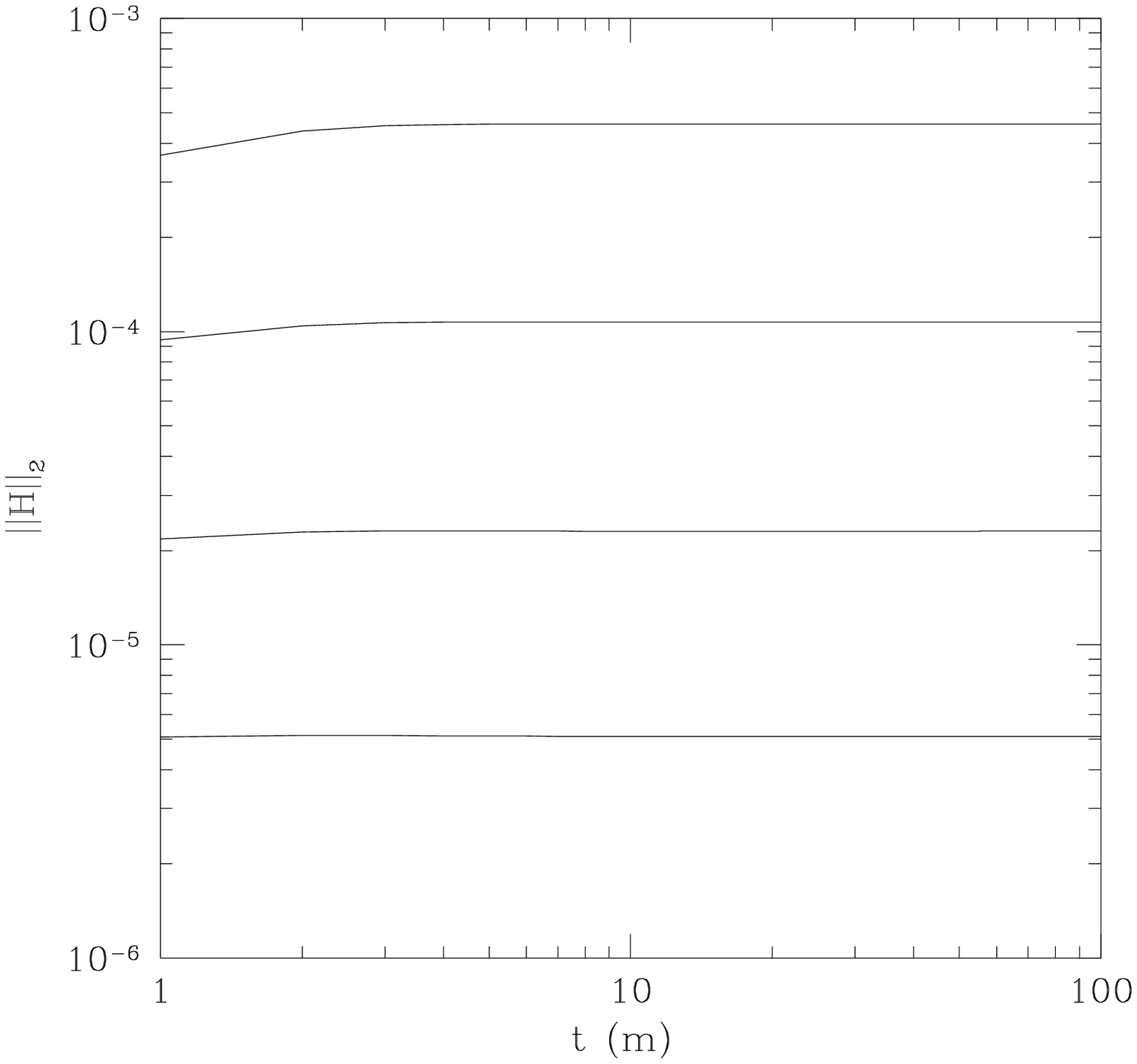}}
\caption{
$L_2$ norm of the Hamiltonian 
constraint as a function of time 
for iEF initial data and EL+AL lapse-shift.
The computational domain extends from $r_e = 1\,m$ to $r_o = 40\,m$. 
Lines from top to bottom correspond to resolutions of
$\Delta r = m/5,\, m/10,\, m/20,\, m/40$, respectively. 
}
\label{fig:elal1}
\end{figure}

Figure~\ref{fig:elal1} shows the $L_2$ norm of the Hamiltonian constraint
as a function of time.
The initial data is given by the iEF analytic solution (\ref{ef}),
and the lapse and shift are chosen from the EL+AL conditions.
Similar results were obtained with PG coordinates. 
The computational domain extends from $r_e = 1\,m$ to $r_o = 40\,m$.
We tried larger and smaller values for $r_o$. However, the stability of the
simulations was not affected by the location of the outer boundary. 
We use an upwind parameter $q = 0.5$ and a constraint-term parameter $\mu = 2$.
We show runs for resolutions of
$\Delta r = m/5,\, m/10,\, m/20,\, m/40$  with 
$\Delta t = 0.25\, \Delta r$. 
The run with $\Delta r=m/5$ has a resolution
similar to those used in three-dimensional simulations 
of black hole collisions.

\begin{figure}
\centerline{\epsfxsize=250pt\epsfbox{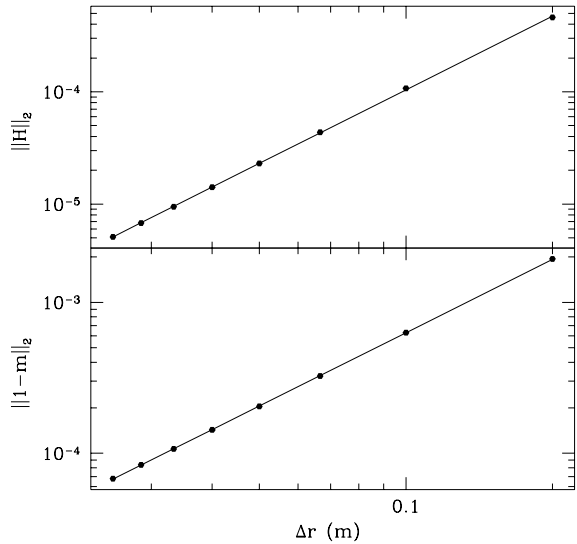}}
\caption{
$L_2$ norm of the Hamiltonian constraint (top) and
$L_2$ norm of the mass error (bottom) as a 
function of resolution. The errors plotted were obtained at $t=200\,m$.
The convergence rate implied by the Hamiltonian errors is 2.18 and
by the mass error 1.7.
}
\label{fig:elal2}
\end{figure}

\begin{figure}
\centerline{\epsfxsize=250pt\epsfbox{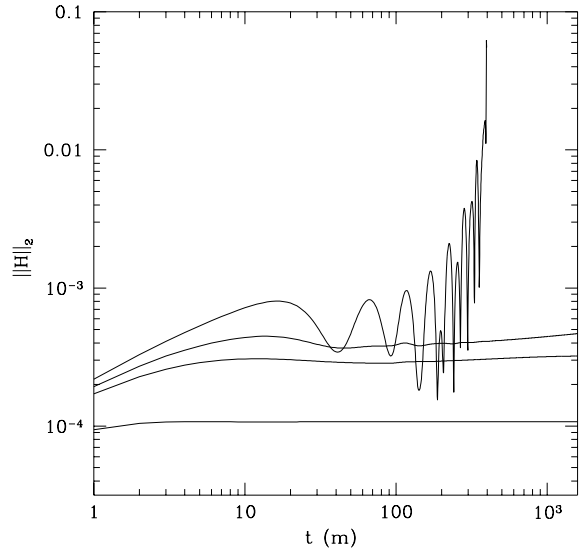}}
\caption{
$L_2$ norm of the Hamiltonian constraint as
a function of time for $\Delta = m/10$ and $r_o = 40\,m$.
Each line corresponds to different values of the
constraint-term parameter $\mu$.
The values of $\mu$ are 0.0, 0.025, 0.5 and 2
in order of stability improvement.
}
\label{fig:elal3}
\end{figure}

Figure~\ref{fig:elal2} shows the $L_2$ norm of the Hamiltonian constraint (top)
and the $L_2$ norm of the mass (\ref{eq:mass_f}) error (bottom)
taken at time $t=200\,m$ as a function
of resolution. 
The convergence rate implied by the Hamiltonian constraint is 
2.18 and by the mass error  1.7.
The convergence rate from the Hamiltonian constraint is larger than
second order because we used third order accurate discretizations of the 
advection term as well as third order accurate extrapolations at the excision. 
On the other hand, the convergence rate obtained 
from the mass error is less than second order
because the mass function is proportional to $g^{rr}$, a quantity difficult to
handle numerically near the singularity.
The reason for using the Hamiltonian constraint and mass to monitor accuracies and
convergence is because of the gauge invariant nature of these quantities.
Finally for reference, we show in Fig.~\ref{fig:elal3} the $L_2$ norm of the solution
Hamiltonian constraint as a function of time for runs with resolution $\Delta r = m/10$ varying
the parameter $\mu$. 
Different lines correspond to values of $\mu = 0.0,\, 0.025,\, 0.5,\,$ and 2
in order of stability improvement.
It is clear from this figure
the dramatic effect that the added constraint term has on the stability of
the simulations.

\subsection{Ingoing Null + Area Locking}

The ingoing null + area locking (IN+AL)
recipe to specify the lapse and shift consists of, in addition to locking
the areal coordinate, imposing the condition that
the vector $\partial _t - \partial _r$ remains null throughout the evolution. 
This null condition is only compatible with the iEF case since by construction
the iEF solution is based on ingoing null observers. An analogous (ingoing timelike) condition
can be obtained for the PG case. 
In terms of the spacetime metric, the ingoing null condition is stated as
$g_{tt}-2\,g_{tr}+g_{rr} = 0$, or similarly in terms of 3+1 metric
functions in (\ref{spherical}) as 
\be
\alpha = a\,(1-\beta)\,.\label{eq:null}
\ee
Conditions (\ref{eq:null}) and (\ref{lock}) yield 
\begin{equation}
\alpha = \frac{a}{arK_b + 1} \;\;\;\hbox{and}\;\;\; 
\beta = \frac{arK_b}{arK_b + 1} \;,   \label{inal_gauge}
\end{equation}
where we have set $b=r$ since by construction $b$ remains locked
to $r$. This prescription for the lapse and shift has been successfully applied in the past
\cite{marsa,jonathan,oscar}.

In order to make the ADM equations in the IN+AL gauge a first order in space
system, we need to introduce two new variables: $w \equiv \partial_r\,a$ 
and $y  \equiv \partial_r\,K_b$.
Even after this choice is made, there is no unique way of writing the
resulting equations as a quasi-linear system. The reason is the ambiguity
one encounters when dealing with the terms involving $\partial_r\,a$. One has
the choice to either keep it as
$\partial_r\,a$, substitute it with $w$, or a combination of both. 
Either choice changes the principal part of
the equation. However, in our case, we need only to find a choice that yields
a strongly hyperbolic system. It turns out that the simplest choice of  
replacing $\partial_r\,a$ by $w$ everywhere 
yields a system that is well-posed.
That is, the resulting ADM equations have the form 
\be
\partial_t\, u = A\,\partial_r\,u + B\,u\,, 
\label{eq:inal}
\ee
with
\be
u = \left(  \begin{array}{c} a  \\ K_a  \\ K_b \\ y \\ w 
\end{array}
\right) \; 
\ee
and
\be
A = \frac{1}{z^2}\,\left( \begin{array}{ccccc}
0 & 0 & 0 & 0 & 0 \\
 & & & & \\
0 & \displaystyle{z\,a\,r\,K_b} & 0 &
    \displaystyle{r} & 
    \displaystyle{-a^{-2}} \\  
 & & & & \\
0 & 0 & 0 & 0 & 0 \\
 & & & & \\
0 & \displaystyle{z\,a\,K_b} & 0 &
    \displaystyle{z+a^2r^2K_b^2} &
    \displaystyle{a^{-1}\,K_b}  \\ 
 & & & & \\
0 & - \displaystyle{z\,a^2} & 0 & 
      \displaystyle{a^2\,r} &
      \displaystyle{a\,r\,K_b(1+z)}
\end{array}
\right)\,,
\ee
where we have introduced $z \equiv 1 + a\,r\,K_b$
to simplify notation.
The corresponding eigenvalues and eigenvectors are
\begin{eqnarray}
\lambda_1 & = & 0 \; \mbox{ with eigenvectors} \;\;\; 
\vec{e}_1 = [1,0,0,0,0] \;\;\; , \;\;\; \vec{e}_2 = [0,0,1,0,0] \; , \label{zero_eg}\\
& & \\
\lambda_2 & = &  1 \; \mbox{ with eigenvectors} \;\;\;
\vec{e}_3= [0,1,0,0,-z\,a^2] \;\;\; , \;\;\; 
\vec{e}_4 = [0,0,0,1,a^2 r] \; , \\ 
& & \\
\lambda_3 & = & \frac{arK_b-1}{z} \mbox{ with eigenvector} \;\;\;
\vec{e}_5= [0,1,0,-aK_b,a^2] \; .
\end{eqnarray}
Notice that all the eigenvectors are independent, and, thus, the system is
strongly hyperbolic. Also notice that the eigenvalues $\lambda_2$ and $\lambda_3$
are again the characteristic speeds ($\beta \pm \alpha/a$) along the light cone. 
Furthermore, in iEF coordinates, $\lambda_3 = (2\,m-r)/(2\,m+r)$.
Therefore, one encounters a similar situation to that of the EL+AL
case; namely, at $r_e$ all the eigenvalues are non-negative,
and at $r_o$ one has that only $\lambda_3 < 0$.
The existence, uniqueness, and well-posedness for IN+AL follows then
as with EL+AL case.

Regarding the constraints, their evolution is also described by a
strongly hyperbolic system with characteristic speeds along the
light cone. It is important to stress that our ADM equations
already imply this, i.e. it does not have any relation to making the
evolution equations first order in space. In fact, the principal part of the
evolution equations for the constraints is
exactly the same as in the EL+AL gauge, but now the lapse
and shift are given by (\ref{inal_gauge}). Thus, the analysis and conclusions
also follow as in that case. 

An interesting aspect of the IN+AL choice is that
it is possible to find the general solution to Einstein's equations.
We start by defining $f \equiv r\,a\,K_b$. 
We then use the momentum constraint (\ref{eq:m_const}) to eliminate $K_a$ from
the other equations. The outcome is that we need only to 
solve three of the four equations
(\ref{adot}), (\ref{kadot}-\ref{eq:h_const}).
We choose to work with the Hamiltonian constraint,
(\ref{eq:h_const}), and Eqs. (\ref{adot}) and (\ref{kbdot}). The resulting
system of equations reads
\ba
0 &=& a\,(f^2+a^2-1)+ 2\,r\,a\,f\,\partial_r f
+ 2\,r\,(1-f^2)\,\partial_r a  \label{ham} \\
(\partial_t - \partial_r)\,a &=& 
\frac{a\,(f^2+a^2-1)}{2r(1+f)^2} \\
(\partial_t - \partial_r)\,f &=& \frac{(f^2+a^2-1)}{r(1+f)}\,.
\ea
The general solution of these equations is
\ba
f   &=& \frac{2\,m}{r} - C\left(1-\frac{2\,m}{r}\right)\\
a^2 &=& (1+C)\,(1+f)
\ea
where $C=C(t+r)$ and $m$ a constant.
In terms of the 3+1 variables, the solutions reads
\be
a^2 = (1+C)(1+f) \;\; , \;\;
b^2 = r^2 \;\; , \;\;
K_a = \partial_r\left(\frac{f}{a}\right) \;\; , \;\;
K_b = \frac{f}{a\, r} \;\; , \;\;
\alpha = \frac{a}{1+f} \;\; , \;\;
\beta = \frac{f}{1+f}\,,
\ee
with the line element (\ref{spherical}) explicitly given by
\ba
\label{keith_sol}
ds^2 & = &  -\,(1+C)^2\,\left(1-\frac{2\,m}{r}\right) \,dt^2\,
+\,2\,(1+C)\,\left[\frac{2\,m}{r}
- \left(1- \frac{2\,m}{r}\right) C\right]\,dt\,dr \\
& & +\,(1+C)\,\left[1+\frac{2\,m}{r}
- \left(1- \frac{2\,m}{r}\right) C\right]\,dr^2\,
+\,r^2\,d\Omega^2\,.
\ea
By setting $C=0$ one recovers the iEF solution (\ref{ef}).
Also, it is not difficult to show from the gauge invariant
definition of mass (\ref{eq:mass_f})
that the parameter $m$ is indeed the mass of the black hole.
An important property of the general solution (\ref{keith_sol}) 
is that it shows explicitly the residual gauge freedom associated 
with the IN+AL choice of lapse and shift. We have found explicitly the 
equivalence-class of solutions gauge related to the iEF solutions that satisfy
the IN+AL lapse-shift condition.

\begin{figure}
\centerline{\epsfxsize=250pt\epsfbox{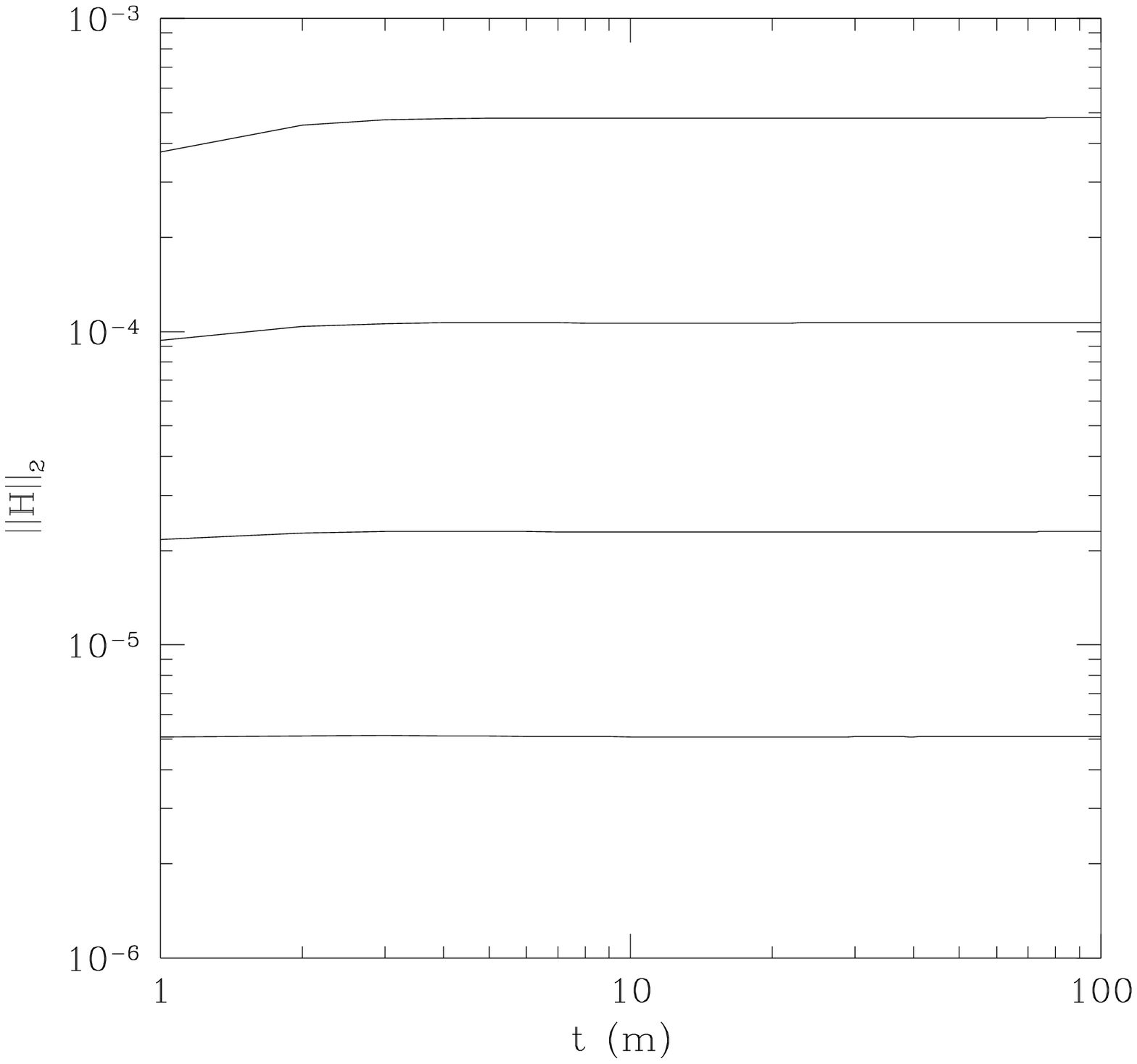}}
\caption{
Same as in Fig.\ref{fig:elal1} but for the IN+AL case.
}
\label{fig:inal1}
\end{figure}

\begin{figure}
\centerline{\epsfxsize=250pt\epsfbox{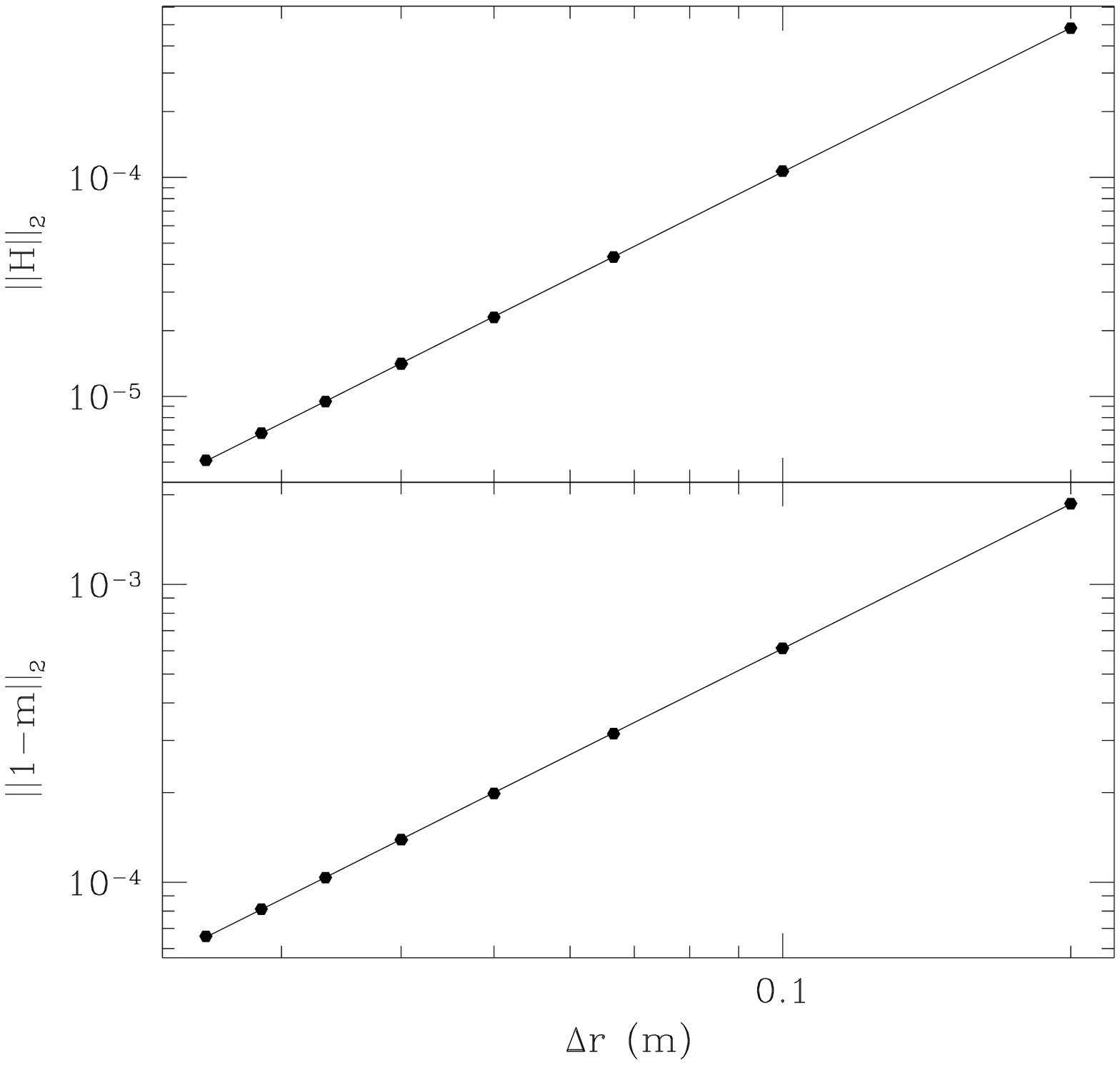}}
\caption{
Same as in Fig.\ref{fig:elal2} but for the IN+AL case.
}
\label{fig:inal2}
\end{figure}

\begin{figure}
\centerline{\epsfxsize=250pt\epsfbox{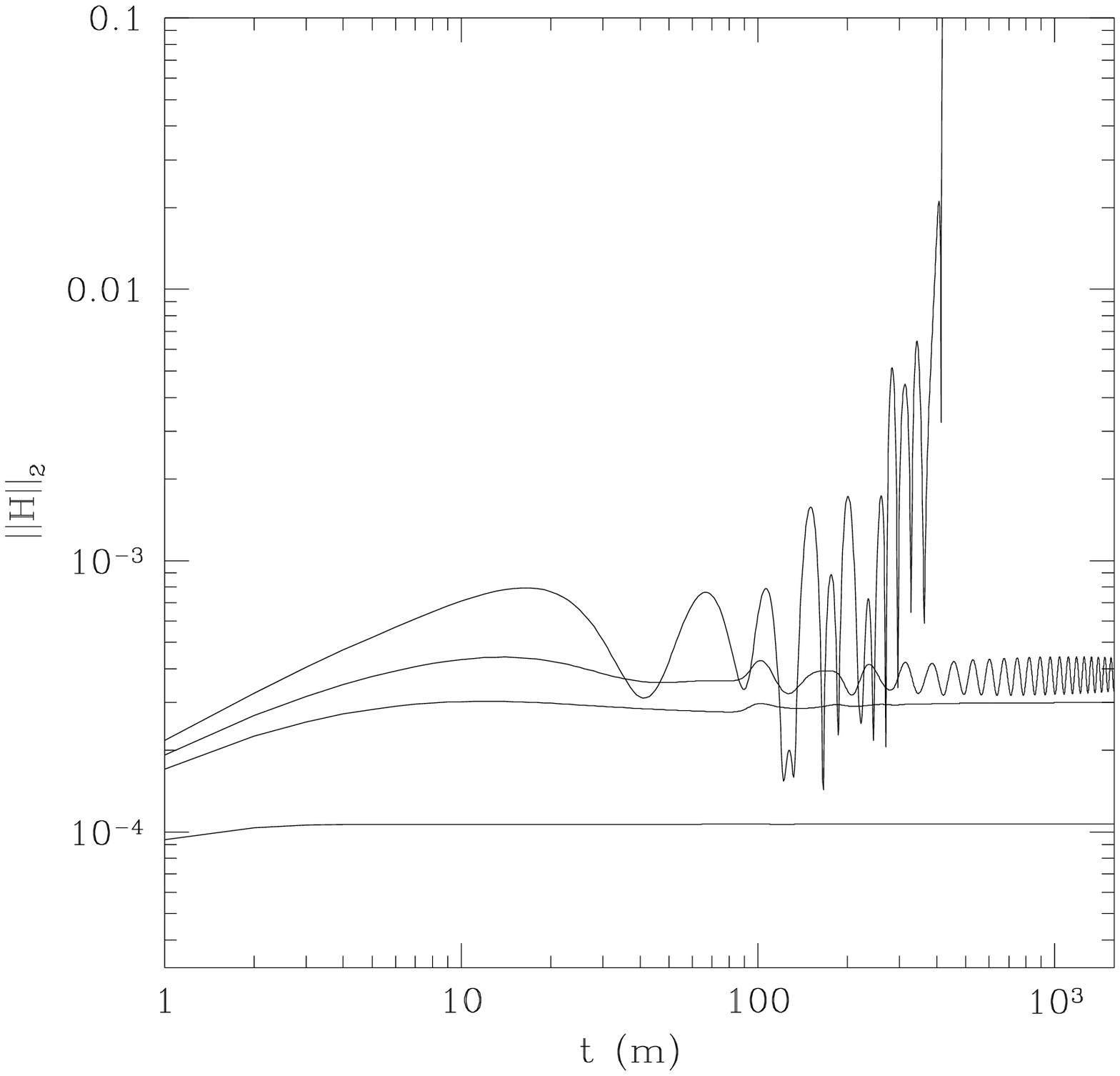}}
\caption{
Same as in Fig.\ref{fig:elal3} but for the IN+AL case.
}
\label{fig:inal3}
\end{figure}

We repeat the same type of numerical experiment as with the EL+AL case,
same parameter values, resolutions, boundary conditions and initial data.
The lapse and shift are however constructed in this case from (\ref{inal_gauge}). 
Figure~\ref{fig:inal1} shows the $L_2$ norm of the Hamiltonian 
constraint as a function of time for different resolutions.
Figure~\ref{fig:inal2} shows the $L_2$ norms of the Hamiltonian constraint (top)
and mass error (bottom) 
for different resolutions. The convergence rate are similar to those
in EL+AL, namely 2.19 from the Hamiltonian constraint
and 1.7 from the mass error.
Finally, Fig.~\ref{fig:inal3} shows the $L_2$ norm of the Hamiltonian constraint
as a function of time for a resolution of $\Delta r = m/10$.
Different lines correspond to values of $\mu = 0.0,\, 0.025,\, 0.5,\,$ and 2
in order of stability improvement.
It is clear from this results that the stability behavior 
of the system of equations under the IN+AL gauge choice follows closely that
of EL+AL.

\subsection{Exact Lapse + Exact Shift} 
 
Finally, we consider the case in which the lapse function and shift vector
are prescribed by the exact analytic solutions.
To date, it has not been possible to obtain with the
standard ADM system long-term stable and convergent numerical evolutions
under the EL+ES choice. Before we present results of EL+ES evolutions
using the adjusted ADM system of equations, let us investigate
its quasi well-posedness properties. 
To make the evolution equations first order in space, we add a new variable
$y=\partial_r b$. Once more, the resulting set of equations have the form
\be
\partial_t\, u = A\,\partial_r\,u + B\,u\,, 
\label{eq:eles}
\ee
with
\be
u = 
\left(
\begin{array}{c}
a \\
b \\
K_a \\
K_b\\
y
\end{array}
\right)
\ee
and
\be
A= 
\left( 
\begin{array}{ccccc}
\beta & 0 & 0 & 0 & 0 \\
& & & & \\
0 & 0 & 0 & 0 & 0 \\
& & & & \\
\displaystyle{\frac{b\,\partial_r \alpha +2\,(1-\mu)\,\alpha y}{a^3b}}
& 0 & \beta & 0 & \displaystyle{-\frac{2\,(1-\mu)\,\alpha}{a^2b}} \\
& & & & \\
\displaystyle{\frac{\alpha y}{ba^3}} & 0 & 0 & \beta &
\displaystyle{-\frac{\alpha }{a^2b}} \\ 
& & & & \\
0 & 0 & 0 & -\alpha b & \beta
\end{array}
\right)\,.
\ee
The matrix $A$ has eigenvalues
\begin{eqnarray}
\lambda _1 &=& \beta \\
\lambda _2 &=& \beta \\
\lambda _3 &=& 0 \\
\lambda _4 &=& \beta+\alpha/a\\
\lambda _5 &=& \beta-\alpha/a\,. 
\end{eqnarray}
There are two fields that propagate with characteristic speed ($\beta$)
along the normal to the hypersurfaces in the foliation, one field with zero speed and
two other fields with characteristic speeds ($\beta\pm\alpha/a$)
along the light cone.
It is not difficult
to show that the eigenvectors corresponding to the
eigenvalue $\lambda_1 = \lambda_2 = \beta$ are not distinct, whatever
the value of $\mu$. Therefore,
the system of equations is only weakly hyperbolic and thus not quasi well-posed, 
as we have already seen by considering the
EL+ES gauge perturbations by themselves in Sec.~\ref{sec:gmodes}. 

As in the previous two cases, the
the constraints propagate according to 
\begin{equation}
\partial_t\,v = P\,\partial_r\, v + Q\, v \label{eles_cons} \; ,
\end{equation}
where now
\be
v =
\left( 
\begin{array}{c}
H \\
M 
\end{array}
\right)
\ee
and
$$
P= 
\left( 
\begin{array}{cc}
\beta & \displaystyle{\frac{4\alpha }{a^2}}  \\
&  \\
\displaystyle{\frac{\alpha }{4} }& \beta
\end{array}
\right)
$$
This is, we have the same situation as before, namely propagation of the constraints
along with characteristic speeds along the light cone.

\begin{figure}
\centerline{\epsfxsize=450pt\epsfbox{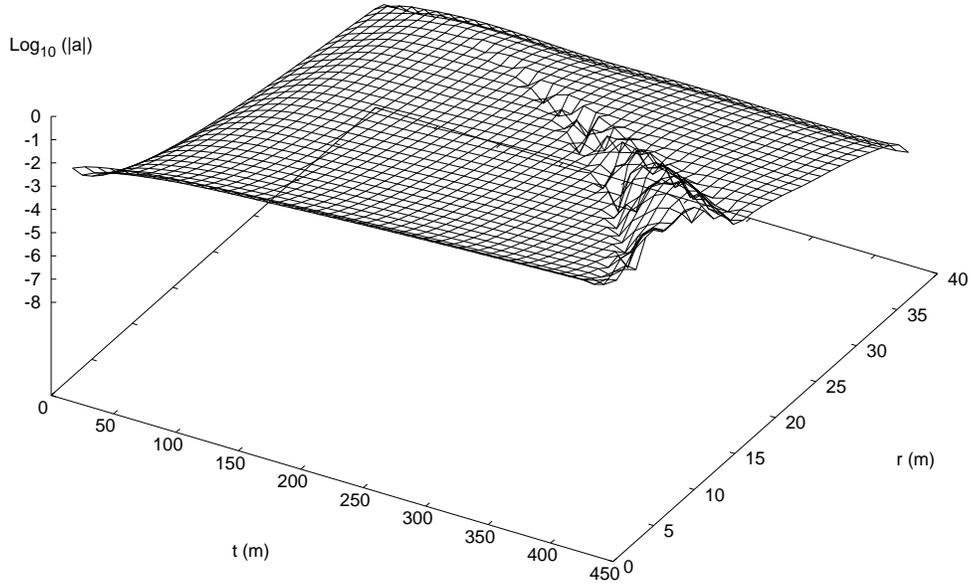}}
\caption{
Solution error for the metric function $a$ in iEF coordinates
for the EL+ES case. The resolution is $\Delta r = m/10$,
and the outer boundary is located at $r_o = 40\,m$. A pulse originated
at the location of the outer boundary due to discontinuities
in the truncation error propagates 
in the direction of the black hole (decreasing $r$-coordinate).
The simulation stops because this pulse-error increases to the point that
the metric function $a$ becomes negative.
}
\label{fig:eles1}
\end{figure}

We have carried out the same numerical experiments as for the previous two cases. 
However, for the iEF solution, the simulations with computational domains 
for $r_o \ge 40\,m$ crashed. 
We have been able to identify two possible sources behind this problem. One of them
is the pulse originated at the outer boundary due to
discontinuities of the truncation error. This pulse
propagates in the direction of the black hole (i.e. decreasing $r$-coordinate)
with the characteristic speed $\beta$. As the pulse
travels its amplitude grows in time. This effect is shown
in Fig.~\ref{fig:eles1}. Here we plot
the solution error for the metric function $a$ as a function of space and time
for a resolution $\Delta r = m/10$
and outer boundary located at $r_o = 40\,m$. 
The simulation stops because this pulse-error increases to the point that
the metric function $a$ becomes negative.
For small computational domains ($r_o < 40\,m$), 
the crossing time of this pulse is short enough and does not
allow a catastrophic growth of the pulse.
The pulse is able to leave the computational domain through
the excision boundary without crashing the simulation. 
Since the initial amplitude of this pulse is
$O(\Delta r^2)$, i.e. the accuracy of the discretization, in principle one could find
a resolution small enough such that the growth of the pulse does not affect
the life of the simulation. 
However, accessing those fine resolutions in three-dimensional
simulations is likely to be impractical.
The second, and perhaps more severe, source of the problem is the presence of
the zero velocity mode. Zero velocity modes were in principle also allowed in the case
of IN+AL, see Eq.~(\ref{zero_eg}).  However, the adjusted IN+AL system yields only numerical
solutions of the general form (\ref{keith_sol}), which clearly does not 
contain a zero velocity mode. If $\mu = 0$, these zero velocity modes are 
not suppressed and eventually terminate the simulation.
This catastrophic effect induced by zero velocity modes has been previously
noticed by Alcubierre et.al \cite{alcub}.

\begin{figure}
\centerline{\epsfxsize=250pt\epsfbox{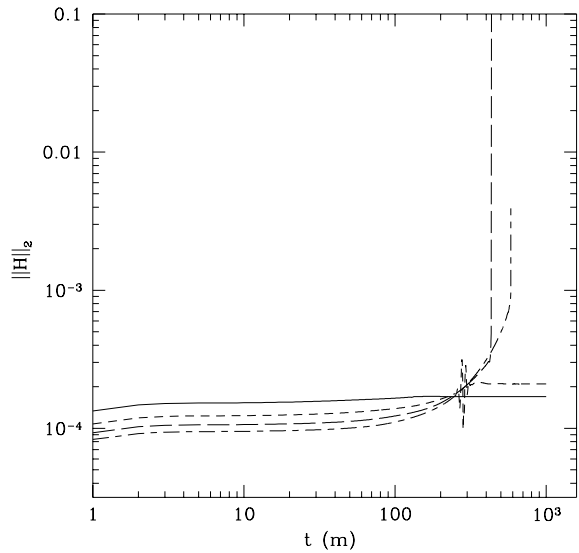}}
\caption{
$L_2$ norm of the Hamiltonian constraint in iEF coordinates
for the EL+ES case as a function of time. 
The resolution is $\Delta r = m/10$ and $\mu = 2$.
Each line corresponds to different location
of the outer boundary: $r_o = 
20\,m\, (\hbox{solid}),\,
30\,m\, (\hbox{short dash}),\,
40\,m\, (\hbox{long dash}),\,
50\,m\, (\hbox{short dash long dash})$.
}
\label{fig:eles2}
\end{figure}

\begin{figure}
\centerline{\epsfxsize=250pt\epsfbox{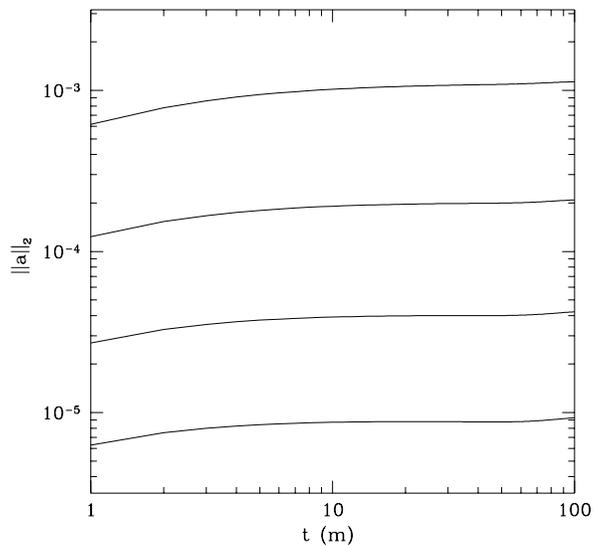}}
\caption{
$L_2$ norm of the solution error in PG coordinates
of the metric function $a$
as a function of time. Each line correspond to different resolutions,
$\Delta r = m/5,\, m/10,\, m/20,\, m/40$, respectively
from top to bottom.
}
\label{fig:eles3}
\end{figure}

Figure~\ref{fig:eles2} shows the $L_2$ norm of the
Hamiltonian constraint as a function of time with resolution $\Delta r = m/10$ for
different locations of the outer boundary. 
For computational domains with approximately $r_o \le 40\,m$,
the zero velocity mode is still present, but  is damped eventually.
The reasons why this mode stops growing remain unclear. 
Nonetheless, there is a strong indication
that this behavior is connected to the particular 
choice of coordinates used to set the exact 
lapse and shift. If one, instead of the iEF solution, sets the lapse and shift
from the PG solution, the outcome of the simulations is completely different.
EL+ES simulations with PG lapse and shift 
are long term stable and convergent for arbitrary sizes
of the computational domain as long as the system of equations was adjusted
with $\mu \ge 1$. Figure~\ref{fig:eles3} shows the $L_2$ norm of the solution error
for the metric function $a$
as a function of time for different resolutions with $r_o = 40\,m$
for EL+ES in PG coordinates.

\section{Conclusions}
\label{sec:end}

We have demonstrated that, at least for the case of single black hole
spacetimes in spherical symmetry, it is possible to obtain
long-term stable and convergent numerical simulations using
the standard ADM system of equations if
the equations are adjusted by introducing terms involving the
constraints. Results were presented for three choices of lapse and
shift. In addition, we introduced the concept of quasi well-posedness,
which appears to be useful in characterizing the properties of the
system of evolution equations. We are currently investigating the
extension of this approach to three-dimensional evolutions.
 
\acknowledgments
This work was supported by 
NSF PHY9800973, NSF PHY0090091 and by the Eberly research funds of 
the Pennsylvania State University.
We wish to thank D.~Arnold, A.~Ashtekar, L.~Lehner, R.~Matzner, R.~Price, O.~Reula,
H. Shinkai and B.~Whiting for comments and helpful discussions.

\eject

\end{document}